\documentclass[conference]{IEEEtran}
\IEEEoverridecommandlockouts
\usepackage{cite}
\usepackage{amsmath,amssymb,amsfonts}
\usepackage{algorithmic}
\usepackage{graphicx}
\usepackage{textcomp}
\usepackage{xcolor}
\def\BibTeX{{\rm B\kern-.05em{\sc i\kern-.025em b}\kern-.08em
    T\kern-.1667em\lower.7ex\hbox{E}\kern-.125emX}}
\begin{document}

\title{Respiratory Rate Estimation from Face Videos
}

\author{Mingliang Chen\textsuperscript{1}, Qiang Zhu\textsuperscript{1}, Harrison Zhang\textsuperscript{2}, Min Wu\textsuperscript{1} and Quanzeng Wang\textsuperscript{3}\\
\textsuperscript{1}University of Maryland, College Park, USA~~~~~~~~~~~~~~\textsuperscript{2}Georgia Institute of Technology, Atlanta, USA\\
\textsuperscript{3}Food and Drug Administration, White Oak, USA\\
\small{\textsuperscript{1}\{mchen126,zhuqiang,minwu\}@umd.edu~~~\textsuperscript{2}hzhang462@gatech.edu~~~\textsuperscript{3}quanzeng.wang@fda.hhs.gov}
}


\maketitle

\ifodd 1
\newcommand{\rev}[1]{{\color{blue}#1}} 
\newcommand{\com}[1]{\textbf{\color{red}(COMMENT: #1)}} 
\else
\newcommand{\rev}[1]{#1}
\newcommand{\com}[1]{}
\fi

\begin{abstract}
Vital signs, such as heart rate (HR), heart rate variability (HRV), respiratory rate (RR), are important indicators for a person's health. Vital signs are traditionally measured with contact sensors, and may be inconvenient and cause discomfort during continuous monitoring. Commercial cameras are promising contact-free sensors, and remote photoplethysmography (rPPG) have been studied to remotely monitor heart rate from face videos. For remote RR measurement, most prior art was based on small periodical motions of chest regions caused by breathing cycles, which are vulnerable to subjects' voluntary movements. This paper explores remote RR measurement based on rPPG obtained from face videos. The paper employs motion compensation, two-phase temporal filtering, and signal pruning to capture signals with high quality. The experimental results demonstrate that the proposed framework can obtain accurate RR results and can provide HR, HRV and RR measurement synergistically in one framework.
\end{abstract}


\section{Introduction}
\label{sec:intro}
Vital signs monitoring is important for clinical diagnostics and in-home health monitoring. Vital signs such as heart rate (HR), HR variability (HRV), and respiratory rate (RR), are usually measured with non-invasive electrocardiography (ECG) or photoplethysmography (PPG) sensors in clinical examination, or with commercial wearable devices in health monitoring. The measurements in both scenarios often employ contact sensors, which may be inconvenient or cause discomfort in long-term monitoring sessions. For example, it is hard to put sensors on young children and ask them to keep still during the monitoring session.

Some pioneering works reveal a possible approach of remote vital signs measurement with contactless sensors. The studies~\cite{verkruysse2008remote,poh2011advancements} extracted HR from face videos, based on the small color change on the face that is consistent with the pulse signal. This technology is called remote PPG (rPPG). The studies in~\cite{li2014remote,lam2015robust,tulyakov2016self,zhu2017fitness} proposed algorithms to deal with motion and illumination variations in realistic resting scenarios or under movement during fitness training. In addition, a skin subspace model was explored in~\cite{wang2017algorithmic} to enhance the signal-to-noise ratio (SNR) of the extracted rPPG.

Compared to remote HR measurement, remote RR measurement has not yet gained as much attention in contact-free health monitoring research. This physiological modality is important, as an abnormal respiratory rate is a predictor of potentially serious clinical events~\cite{cretikos2008respiratory}. In this paper, we focus on remote RR measurement. There have been several studies exploring remote RR measurement from various sources:
\begin{itemize}
    \item Periodical motions on chest region. Due to the volume change of the lungs during breathing cycles, motion analysis around chest region in video sequences \cite{benetazzo2014respiratory,tveit2016motion} were proposed for remote RR measurement from color or depth videos. In addition, radio frequency signal~\cite{yang2017vital} is another tool to measure the tiny movement around the chest area, similar to RF ranging. However, these methods inevitably are motion sensitive and vulnerable to subjects’ voluntary movements.
    \item Thermal imaging. Infrared cameras can capture the temperature variation around the nostril areas~\cite{pereira2015remote}, which is caused by nasal airflow during inspiration and expiration. However, this method requires special equipment which can be expensive currently for wide deployment, especially for home care.
\end{itemize}

Reference~\cite{verkruysse2008remote,poh2011advancements,li2014remote,lam2015robust,tulyakov2016self,zhu2017fitness} have demonstrated that color cameras are promising contactless sensors for measurement of HR-related signals, i.e. rPPG. HRV has also been shown to be a benign and natural phenomenon in heartbeats that is influenced by breathing~\cite{saykrs1973analysis}. A person's heart rate tends to increase when he/she breathes in, and fall when he/she exhales. While the studies in~\cite{dash2010estimation,karlen2013multiparameter} have demonstrated that ECG and PPG signals can be good sources to measure HRV and RR, few research has been conducted to extract RR systematically from rPPG signals.

In this paper, we establish a systematic framework to measure RR from face videos. We employ motion compensation, a two-phase temporal filtering, and signal pruning to capture high-quality HRV and RR signal. In addition, the proposed framework incorporates modules of multiple vital signal measurements, including HR, HRV and RR. Experimental results show the feasibility and effectiveness of remote RR measurement from face videos.

\section{framework}
\label{sec:framework}

Our proposed framework, as illustrated in Fig.~\ref{fig:roi}, provides simultaneous tracking of three basic and important vital signs, namely, HR, HRV and RR. The framework is composed of the following four main steps. First, we track the region-of-interest (ROI) on the face in the video sequence, and extract the pulse signal, i.e., the rPPG signal. Second, temporal filtering is applied to the rPPG signal to exclude the energy out of the typical HR range. We compute the inter-beat interval (IBI) from the filtered rPPG signal and then obtain HRV. Since HRV obtained based on rPPG is vulnerable to noise contamination caused by such factors as motion, illumination variations, we prune the obtained HRV samples and discard outliers before extracting RR. Thus, in the third step, we model HRV samples with the assumption of a Gaussian distribution and remove the outliers in HRV samples. Finally, spectral analysis is applied on HRV to estimate RR. The details of each step are explained in the following subsections.

\subsection{Rigid Motion Compensation and rPPG Extraction}
\begin{figure}[t]
\centering
\includegraphics[width=1.0\linewidth]{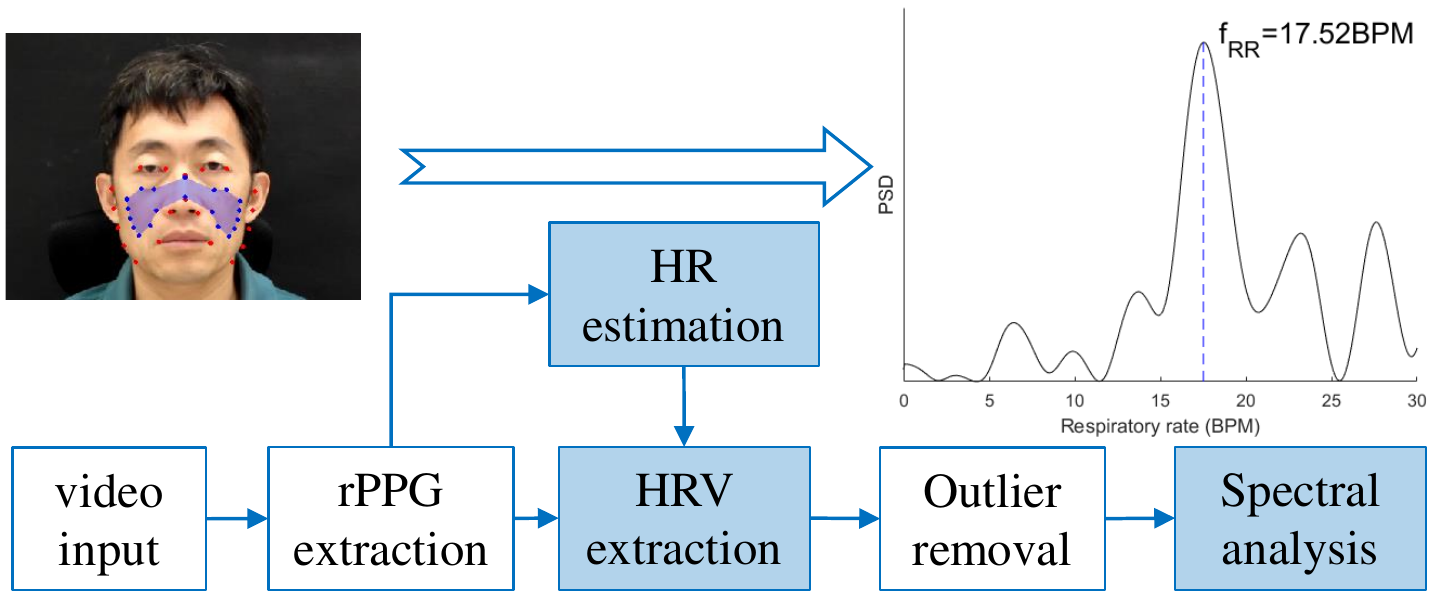}
\caption{The proposed framework for RR measurement from face
videos. The modules in highlight are related to vital sign extraction. The ROI for rPPG extraction is defined in the left-top video frame. The red points are 20 facial landmarks used to define the ROI, which is indicated by the light-blue region.}
\label{fig:roi}
\end{figure}

The aim of rigid motion compensation is to alleviate the noise introduced by the subject's voluntary movement during rPPG extraction. We use the Tasks-Constrained Deep Convolutional Network (TCDCN)~\cite{zhang2016learning} to detect 68 facial landmarks and define the ROI. Fig.~\ref{fig:roi} shows an example of the ROI with the location of 20 facial landmarks. Then, a tracking process is employed to automatically select the corresponding ROI in each video frame. In the tracking process, feature points are first detected inside the face region using the ``good feature to track'' algorithm~\cite{shi1994good}, and then tracked through the following frames using the Kanade-Lucas-Tomasi (KLT) algorithm~\cite{tomasi1991detection}. We denote the locations of $k$ tracked feature points as ${\bf P}_i=[p_i^{1}, p_i^{2}, ..., p_i^{k}]$ and the locations of $l$ ROI boundary points as ${\bf Q}_i=[q_i^{1}, q_i^{2}, ..., q_i^{l}]$ in the $i$-th frame. We estimate the 2D geometric transformation $A$ of the face between two consecutive frames by minimizing the least square error:
\begin{equation}
    \min_{A} \left \| {\bf P}_{i+1}-A{\bf P}_i \right \|^2.
\end{equation}
We then obtain the location of ROI in the next frame from the current frame by ${\bf Q}_{i+1} = A{\bf Q}_i$. Finally, the rPPG signal is extracted by averaging the pixel intensity in the tracked ROI, and is processed with Plane-Orthogonal-to-Skin (POS) algorithm~\cite{wang2017algorithmic} for signal detrending, normalization, and denoising.

\subsection{HRV Extraction}
To capture the fluctuation of HR, we estimate the HR curve and IBIs from rPPG signal. Since the previous work in \cite{li2014remote,lam2015robust,tulyakov2016self} has studied the robust HR measurement from face videos, here we will not discuss HR measurement in detail. Unlike the former studies\cite{li2014remote,lam2015robust,tulyakov2016self} that provide only one heart rate value for a video segment, we track HR throughout the duration of the video via the Adaptive Multi-Trace Carving (AMTC)~\cite{zhu2018amtc}, and obtain the HR curve. AMTC is a robust algorithm for tracking weak signals in severe noise and distortion.

\begin{figure}[t]
\centering
\includegraphics[width=0.9\linewidth]{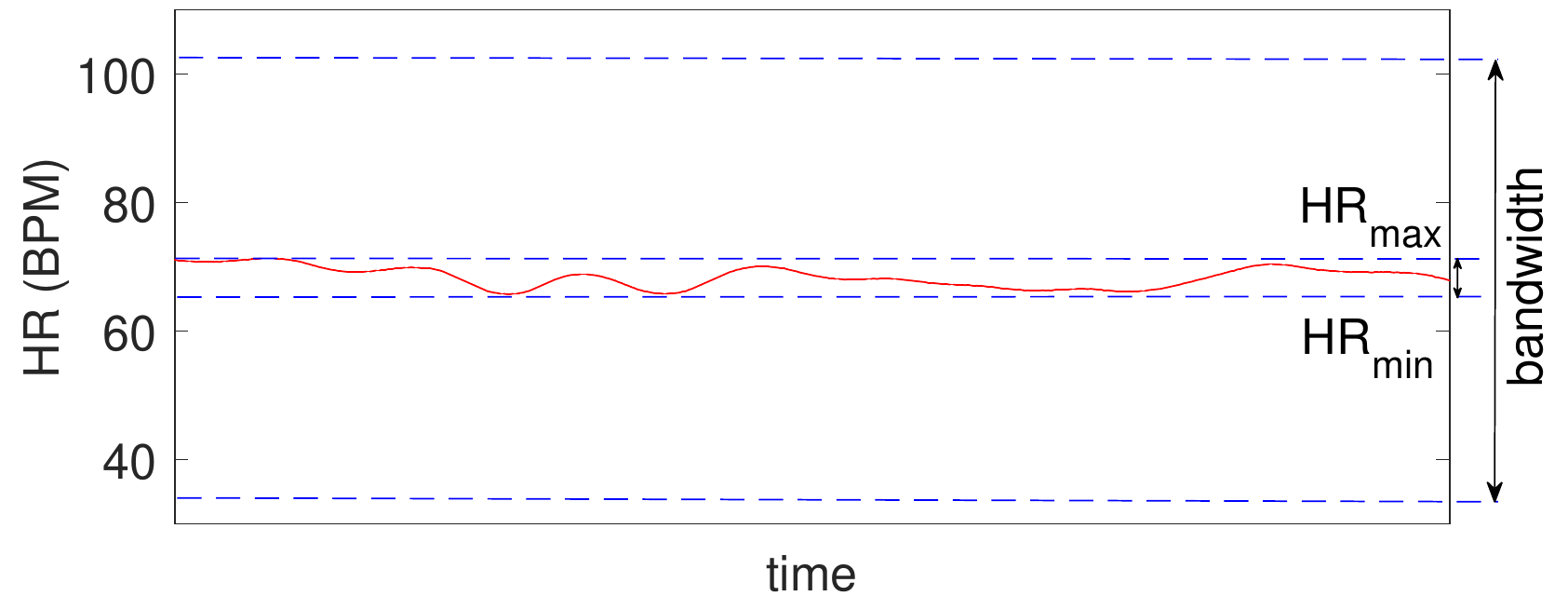}
\caption{The choice of the bandpass of the second filter. The red curve indicates the heart rate curve. The selected bandwidth of the second filter is wider than the dynamic range of HR curve in the video segment.}
\label{fig:filter}
\end{figure}

Note that a temporal bandpass filtering was applied in the former studies~\cite{verkruysse2008remote,poh2011advancements,li2014remote,lam2015robust,tulyakov2016self} to confine the frequency of the rPPG signal within HR range, (i.e., from 0.7~Hz to 4~Hz) to facilitate HR extraction. Although the filtered signal can help perform robust HR estimation, it may not be good enough for HRV extraction, since HRV extraction generally requires the full time domain waveform with high signal quality and clean peaks, and is much more challenging than HR estimation. Therefore, we employ a second temporal filter, an infinite impulse response (IIR) filter with narrower bandpass, and apply it to the rPPG signal to enhance the signal quality. Given the dynamic range of HR from $hr_1$ to $hr_2$, we choose the bandpass for the second filter from $hr_1-${\it offset} to $hr_2+${\it offset}, where {\it offset} is an empirical parameter and should be chosen to exclude as much noise as possible but still preserve the frequency components of HRV. In this paper, We set the {\it offset} to 30 breaths per minute (BrPM), to balance the inclusion of HRV components and the exclusion of noise. Fig.~\ref{fig:filter} illustrates the bandpass of the second filter. To avoid phase distortion, we apply zero-phase filtering in the second filtering, which filters the input signal forward first and then filters the output backward again. The aim of the process is to preserve the location of beats after the filtering and facilitate HRV extraction.

After the second filtering, the peaks in the filtered rPPG signal are detected. Unlike the relatively high sampling frequency in PPG and ECG (usually more than 100~Hz), the frame rate of videos is normally around 30~Hz, which limits the accuracy of the peak locations. We refine the peak locations by interpolating quadratically around the peaks. After the refinement, IBIs are computed from adjacent peaks, with the unit of second. We set the time of the IBI sample as the middle time of the corresponding consecutive peaks. HRV is the reciprocal of IBI: $HRV = \frac{60}{IBI}$, with the unit of beats per minute (BPM). We subtract the HRV signal with the HR curve to extract the fluctuation, and refer to it as detrended HRV in the following.

\begin{figure}[t]
\centering
\includegraphics[width=1.0\linewidth]{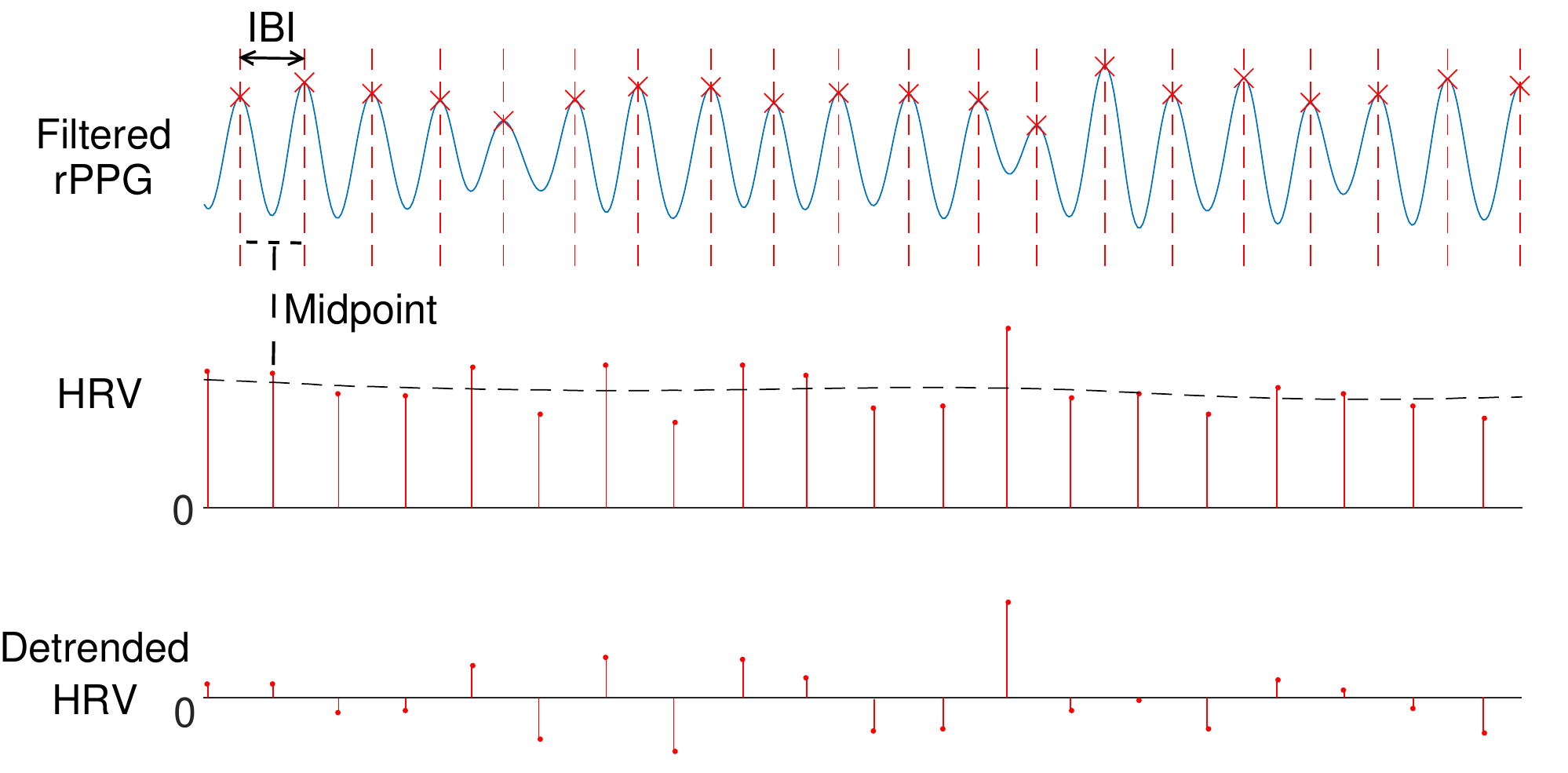}
\caption{Detrended HRV extraction. First, peaks are detected in filtered rPPG signal. After that, IBI signal is extracted and transformed to HRV signal. By subtracting the HR curve (black dash) from HRV signal (red), the detrended HRV is obtained finally.}
\label{fig:ibi}
\end{figure}

\subsection{Outlier Removal}
\begin{figure}[t]
\centering
\includegraphics[width=0.88\linewidth]{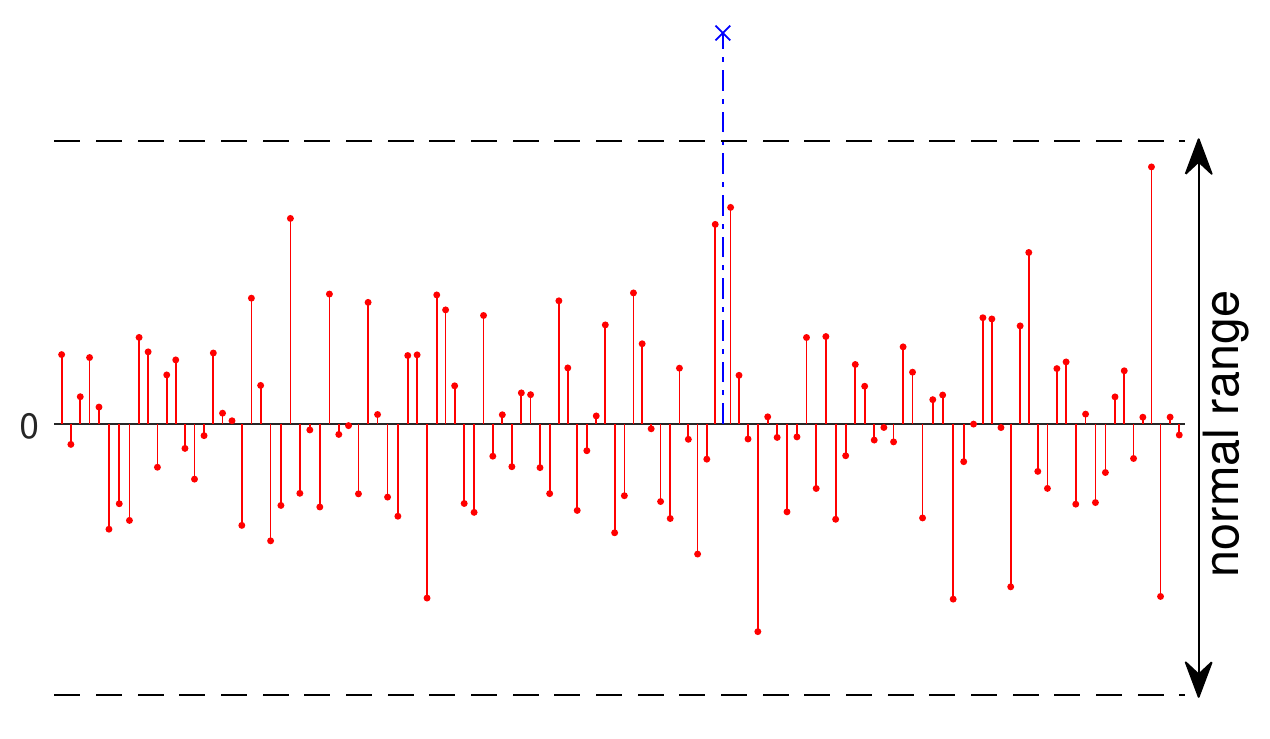}
\caption{Outlier removal. The dash lines are the normal range bounds of the detrend HRV and the blue crossing samples are the excluded outliers.}
\label{fig:outlier}
\end{figure}

Although the second filtering helps us obtain HRV signal in better quality, the signal may still contain some outliers due to noise and error in peak detection. Therefore, the aim of module ``outlier removal'' is to eliminate the bad detrended HRV samples. We examine the detrended HRV samples and eliminate the possible outliers. We first model the detrended HRV samples with Gaussian distribution $\mathcal{N}(\mu,\sigma^2)$, where $\mu$ and $\sigma^2$ can be estimated by maximum likelihood estimation (MLE). Then, we remove the sample $s$ if it is outside the range $[\mu-\alpha\sigma, \mu+\alpha\sigma]$, where $\alpha$ is a hyperparameter. In this paper, we choose $\alpha=3$, according to the ``three-sigma rule''. Fig.~\ref{fig:outlier} illustrates the process of outlier removal.

\subsection{Spectral Analysis}
One of the primary fluctuation sources in HR is respiratory sinus arrhythmia, which is associated with breathing cycles. Hence, employing spectral analysis on detrended HRV can estimate RR. As the obtained detrended HRV sequence is unevenly sampled, we estimate the power spectral density (PSD) of detrended HRV using the Lomb-Scargle periodogram~\cite{lomb1976least,scargle1982studies}. The Lomb-Scargle periodogram is a spectral analysis tool dealing with unevenly sampled data, and is based on a least-squares fit of sinusoids to the data samples. The respiratory rate $f_{RR}$ is obtained by selecting the frequency with the maximal energy response in normal RR range of $5$ to $30$ BrPM.

\section{Experiments}
\label{sec:result}

We evaluated on a set of self-collected face videos to demonstrate the feasibility of remote RR measurement. The dataset contains 60 color video segments from 6 subjects, and each segment lasts around 30 seconds. Each video segment captured the subject's frontal face using a Canon camera affixed on a tripod in a well-lit laboratory. During the data collection, the subjects were asked to sit in front of the camera in rest, but were allowed to have small motions and expressions. In the meantime, the subjects wore a PPG sensor on the fingertip and a respiration belt on the chest, and PPG signals and breathing signals were collected at sampling rate 100~Hz. The breathing signals from respiratory belt were processed by spectral analysis with periodogram to extract RR as the ground truth (GT). PPG-derived RR is obtained using \cite{karlen2013multiparameter} as an anchor reference.

Using the proposed framework, we obtained RR from each video segment and compare it with the corresponding GT from the respiration belt. Bland-Altman plot~\cite{bland1999measuring} is used to evaluate the video-based RR measurement. To demonstrate the effectiveness of the {\it interpolation} in peak finding and {\it outlier removal} in the proposed framework, we also conducted experiments with different module combinations. The two modules, for simplicity, are denoted as {\it Interp} and {\it OR}, respectively.

\begin{figure}[t]
    \begin{minipage}[b]{1.0\linewidth}
      \centering
      \centerline{\includegraphics[width=4.5cm]{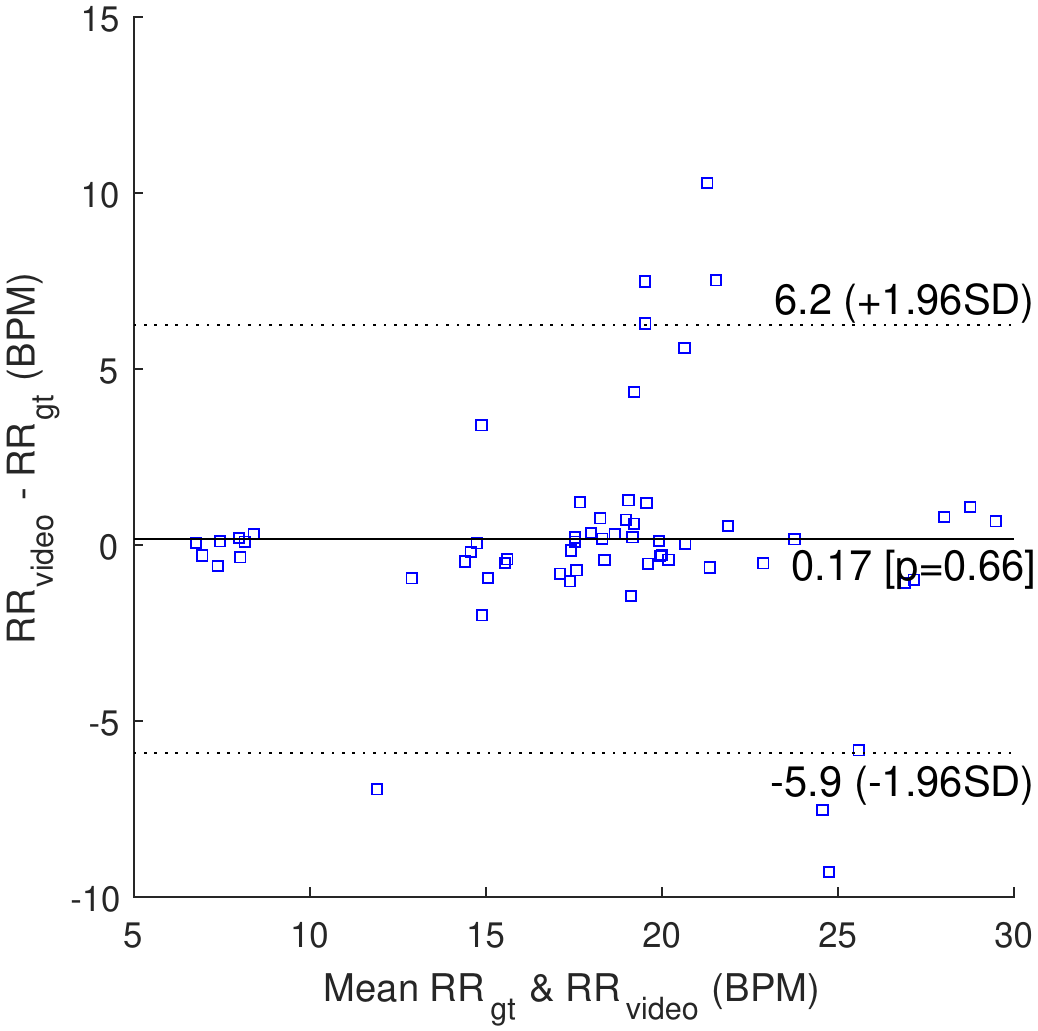}}
      \centerline{\footnotesize{(a) No {\it Interp}, no {\it OR}}}\medskip
    \end{minipage}
    \begin{minipage}[b]{.48\linewidth}
      \centering
      \centerline{\includegraphics[width=4.5cm]{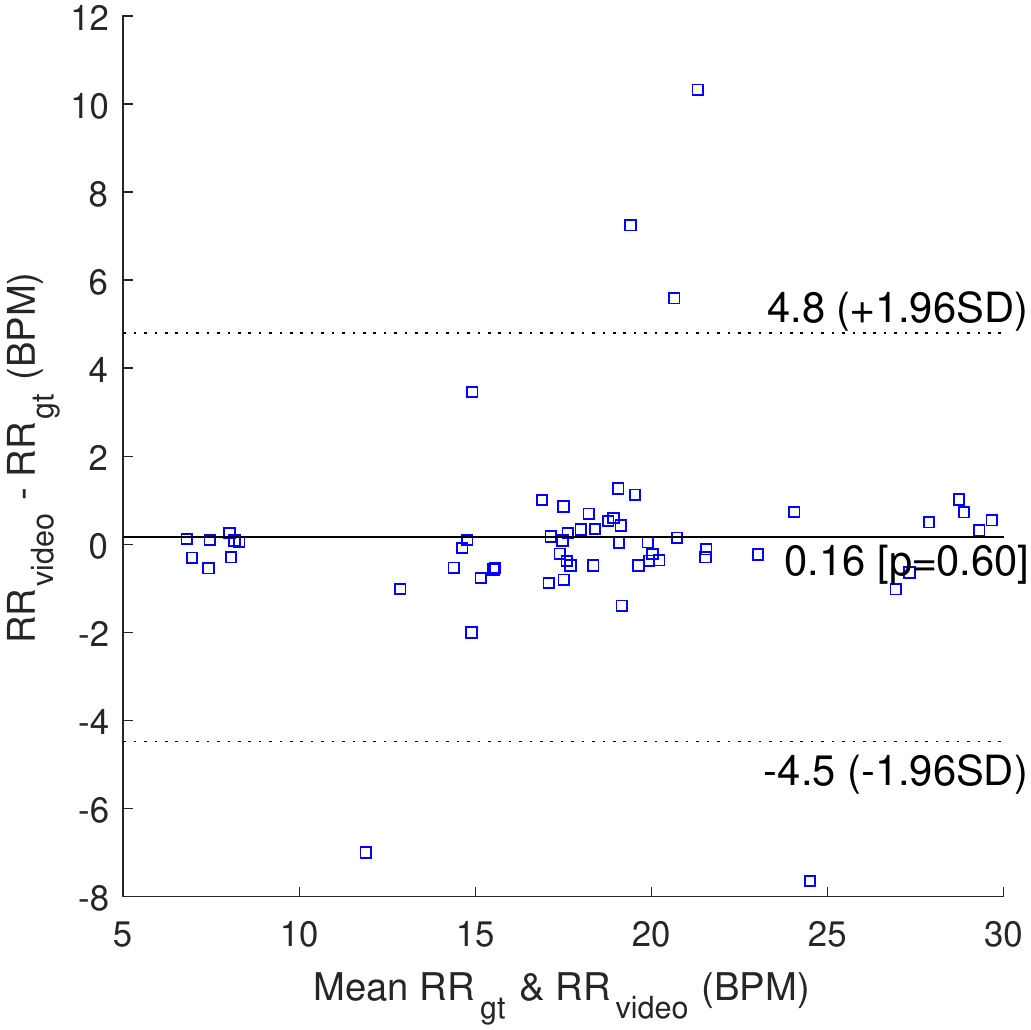}}
      \centerline{\footnotesize{(b) Only {\it Interp}, no {\it OR}}}\medskip
    \end{minipage}
    \hfill
    \begin{minipage}[b]{0.48\linewidth}
      \centering
      \centerline{\includegraphics[width=4.5cm]{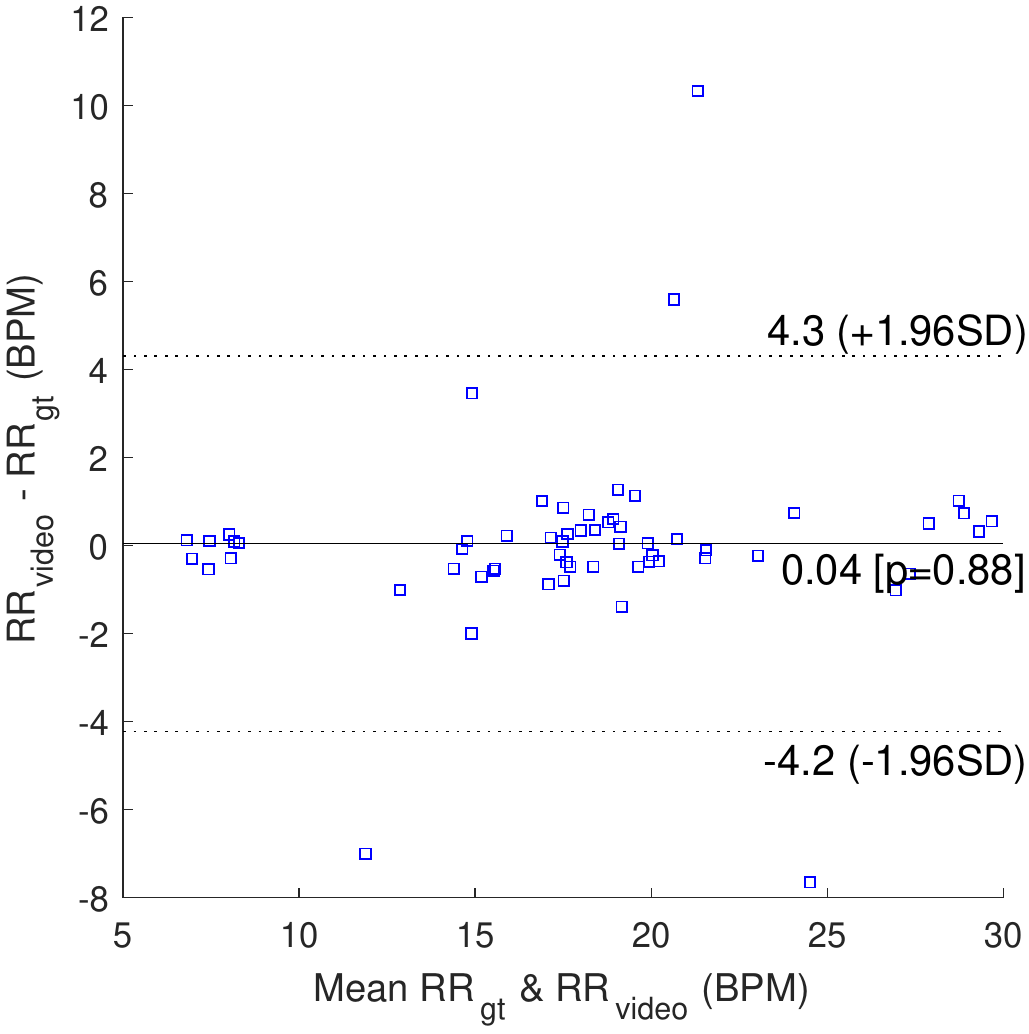}}
      \centerline{\footnotesize{(c) {\it Interp}+{\it OR} (proposed)}}\medskip
    \end{minipage}
    \caption{Bland-Altman plots of the proposed method with different module combinations vs. the respiration belt (GT).}
    \label{fig:ba}
\end{figure}

To evaluate the performance of the proposed method, we consider the metrics in remote HR analysis~\cite{li2014remote,tulyakov2016self}. Specifically, we define the measurement error $RR_e=RR_m-RR_{gt}$,
i.e., the difference between the measured RR $RR_m$ and the ground truth RR from respiration belt $RR_{gt}$. The respiration belt measures the variation of the chest volume with a pressure sensor. As shown in Table~\ref{tab:result}, we report the mean $M_e$ and the standard deviation $SD_e$ of $RR_e$ among all sequences, root mean squared error (RMSE), the mean of error-rate percentage $M_{eRate}=\frac{1}{N}\sum_{i=1}^{N}\frac{\left | RR_e(i) \right |}{RR_{gt}(i)}$, where $N$ is the number of the video sequences and $i$ is the video index. In respiratory rate monitoring, we consider an absolute error of less than 1~BrPM is acceptable. Hence, we present the percentage of estimation where the absolute error was less than 1~BrPM ($\%<1$), as an indicator of the successful rate of RR estimation from videos. Fig.~\ref{fig:ba} presents Bland-Altman plots for conformance testing of the measurement vs. GT obtained from respiration belt.

\begin{table}[t]
\centering
\caption{Performance of the proposed framework. (Unit: BrPM)}
\begin{center}
\begin{tabular}{|c|c|c|c|c|}
\hline
Method      & $M_{e}~(SD_{e})$ & RMSE & $M_{eRate}$ & \%$<1$\\ \hline\hline
PPG-derived (anchor) &  -0.04 (0.15) & 0.15 & 0.77\%  & 100\%    \\ \hline
No {\it Interp}, no {\it OR} & 0.17 (3.10) & 3.08 & 9.30\% & 68.33\%\\ \hline
Only {\it Interp}, no {\it OR} & 0.16 (2.37) &  2.35 & 6.67\% & 76.67\%    \\ \hline
{\it Interp}+{\it OR} (proposed) & 0.04 (2.18) & 2.16 & 5.92\%  & 78.33\% \\ \hline
\end{tabular}
\end{center}
\label{tab:result}
\end{table}

From the results in Table~\ref{tab:result}, we see that PPG-derived RR is an accurate approach to measure RR from heartbeat-related signals, and it also indicates the feasibility of extracting RR from these test subjects using heartbeat-related signal. As shown in Fig.~\ref{fig:ba}, the proposed frameworks with different module combinations can all measure RR accurately in most cases. However, {\it Interp} and {\it OR} modules can enhance the robustness of remote RR measurement, reducing the number of large error cases outside $95\%$ confident interval. From Table~\ref{tab:result}, we can also see {\it Interp} and {\it OR} modules help decrease $M_e (SD_e)$ from $0.17\pm3.10$ to $0.04\pm2.18$ BrPM, $RR_e$ from $3.08$ to $2.16$ BrPM and $M_{eRate}$ from $9.30\%$ to $5.92\%$. The successful rate of estimation also increases from $68.33\%$ to $78.33\%$. Therefore, we include these two modules in our overall framework. As a whole, the results demonstrate the feasibility of remote RR measurement from face videos.

\section{Conclusion}
\label{sec:conclusion}
In this paper, we have developed a systematic framework for remote RR monitoring using face videos. We apply motion compensation, a two-phase filtering, and signal pruning to enhance the robustness of RR measurement. We incorporate HR, HRV, and RR measurement into the proposed framework, aiming to construct a comprehensive system for remote vital signs measurement. Experimental results show that our proposed method is feasible and effective in measuring RR remotely in rest case. Poor signal quality in pulse signal extraction is still a challenge in remote vital sign extraction. How to alleviate the negative influence of voluntary motions and expressions, and illumination changes on pulse signal extraction from the videos is a direction of the future work.

\vspace{2mm} \noindent {\bf Acknowledgement}: The authors would like to thank Dr. Christopher Scully from FDA, for helping obtain PPG and respiration belt data.

\small{\vspace{2mm} \noindent {\bf Disclaimer}: The mention of commercial products, their sources, or their use in connection with material reported herein is not to be construed as either an actual or implied endorsement of such products by the U.S Department of Health and Human Services.}

\bibliographystyle{IEEEbib}
\bibliography{biblist}

\end{document}